\magnification 1200
\centerline  {{\bf Hyperbolic Flows and the Question of Quantum 
Chaos}\footnote*{Based on a lecture in the G.G.Emch Memorial Session 
of the XXXIV Workshop on Geometric Methods in Physics, held in 
Bialowieza, Poland over the period 28 June to 4July, 2015}} 
\vskip 0.5cm
\centerline  {{\bf by Geoffrey L. Sewell}\footnote{**}{e-mail address: 
g.l.sewell@qmul.ac.uk}}
\vskip 0.5cm
\centerline  {\bf Department of Physics, Queen Mary University of 
London,}
\vskip 0.3cm 
\centerline {\bf Mile End Road, London E1 4NS, UK}
\vskip 1cm
\centerline {\bf To the Memory of Gerard Emch}
\vskip 1cm
\centerline {\bf Abstract}
\vskip 0.3cm
Hyperbolic flows, as formulated by Anosov, are the prototypes of chaotic 
evolutions in classical dynamical systems. Here we provide a concise 
updated account of their quantum counterparts originally formulated by 
Emch, Narnhofer, Thirring and Sewell within the operator algebraic 
setting of quantum theory; and we discuss their bearing on the question of 
quantum chaos.
\vfill\eject
\centerline {\bf 1. Introduction}
\vskip 0.3cm
Classical hyperbolic flows, as formulated by Anosov [1], are flows over 
smooth compact connected Riemannian manifolds that admit stable 
expanding and contracting foliations. Thus they are prototype examples 
of chaotic dynamical systems, in that orbits stemming from neighbouring 
points of their phase spaces diverge, generically, exponentially fast from 
one another. 
\vskip 0.2cm
In view of the fundamental character of both quantum ergodic theory 
[2]-[4] and quantum chaology [5, 6], it is natural to ask whether a 
formulation of a quantum counterpart of these flows is feasible. This 
question was addressed by Emch et al  [7] in a treatment that overcame 
the obstacle imposed by the fact that quantum mechanics does not 
accommodate the  differential geometric structures on which the classical 
treatment was based [1, 8].  In fact, their treatment was carried out within 
the framework of operator algebraic quantum theory [9, 10], wherein the 
observables of a model were represented by the self-adjoint elements of a 
$W^{\star}$-algebra and the non-commutative differential structure was 
carried by derivations of that algebra.
\vskip 0.2cm
The present article is devoted to a concise updated account of the picture 
of quantum hyperbolic flows presented in Ref. [7]. Its essential content 
comprises a general formulation of these flows and their chaotic 
properties, together with concrete examples both of models for which 
chaos survives quantisation and models for which it does not.  
\vskip 0.2cm
We start, in Section 2, with a brief account of the classical picture of 
hyperbolic flows. Here the generic model comprises a one-parameter 
group of diffeomorphisms of a manifold that satisfies a certain 
hyperbolicity condition.  Prototype examples of these flows, which we 
provide, are the Arnold cat model and the geodesic flow over a compact 
Riemannian manifold of constant negative curvature.
\vskip 0.2cm
In Section 3 we recast the classical model into the operator algebraic form 
given by the Gelfand isomorphism. This enables us to express the 
hyperbolicity condition in terms of automorphisms of the resultant 
commutative algebra of observables.
\vskip 0.2cm
In Section 4 we provide a simple passage from the classical commutative 
algebraic picture to the quantum non-commutative one, thereby 
formulatng the hyperbolicity condition for the quantum model in term of 
automorphisms of its algebra of observables. In particular we show that 
this condition implies the chaoticity of the quantum model in that the 
evolutes of neighbouring states, as represented by density matrices, 
diverge exponentially fast from one another.
\vskip 0.2cm
In Section 5 we provide an explicit treatment of the quantum version of 
the Arnold cat model and prove that its hyperbolicity, and thus its chaotic 
property, survives the quantisation.    
\vskip 0.2cm
Correspondingly, in Section 6 we provide an explicit treatment of the 
quantum version of the geodesic flow over a compact Riemannian 
manifold of negative curvature and show that, by contrast with the Arnold 
cat model,  it violates the hyperbolicity condition. In other words, 
quantisation of its original classical version destroys its hyperbolicity.
\vskip 0.2cm 
In Section 7 we generalise this result to arbitrary finite quantum 
Hamiltonian systems by showing that they cannot support hyperbolic 
flows.
\vskip 0.2cm
We conclude in Section 8 with a brief discussion of the results presented 
here and their consequences for quantum chaology.
\vskip 0.2cm
The Appendix is devoted to the proof of a key Proposition involved in the 
formulation of the classical hyperbolicity condition of Section 2.
\vskip 0.5cm
\centerline {\bf 2. The Classical Picture}
\vskip 0.3cm
The classical model, ${\Sigma}_{cl}$, is given by a triple $(M, 
{\mu},{\phi})$ [8], where $M$ is a smooth, connected, compact 
Riemannian manifold, ${\phi}$ is a representation of ${\bf R}$ or ${\bf 
Z}$ in the diffeomorphisms of $M$, the representation being continuous 
in the former case; and ${\mu}$ is a ${\phi}$-invariant probability 
measure on $M$. Thus ${\phi}$ and ${\mu}$ represent the dynamics and 
a stationary state, respectively, of the model. Specifically, for $m 
{\in}M$ and $t{\in}{\bf R} \ {\rm or} \ {\bf  Z}, \ {\phi}_{t}m$ is the 
evolute of $m$ at time $t$; and for measurable regions  $A$ of $M, \ 
{\mu}({\phi}_{t}A)={\mu}(A)$. We denote the tangent space at the 
point $m$ of $M$ by $T(m)$ and note that, for fixed time $t$, the 
differential $d{\phi}_{t}$ of  ${\phi}_{t}$ maps $T(m)$ into 
$T({\phi}_{t}m)$. 
\vskip 0.2cm
In order to formulate the condition for the hyperbolicity of the dynamics 
of ${\Sigma}_{cl}$ we first assume that $M$ is equipped with vector 
fields $V_{1},. \ .,V_{n}$, where $n={\rm dim}(M) $ or $\bigl({\rm 
dim}(M) -1\bigr)$ according to whether the time variable $t$ is discrete 
or continuous\footnote*{The difference between $n$ and ${\rm 
dim}(M)$ in the continuous case corresponds to the one dimensionality 
of the space generated by the velocity vector}. It is assumed that at each 
point $m$ of $M$  these fields are linearly independent and that each 
$V_{j}$ has a global integral curve 
$C_{j}(m)={\lbrace}m_{j}(s){\vert}s{\in}{\bf R}; \ 
m_{j}(0)=m{\rbrace}$, given by the unique solution of the equation
$$m_{j}^{\prime}(s)=V_{j}\bigl(m_{j}(s)\bigr); \ 
m_{j}(0)=m.\eqno(2.1)$$  
Thus, the curves ${\lbrace}C_{j}(m){\vert}m{\in}M{\rbrace}$ are 
generated by the action on $M$ of a one-parameter group 
${\lbrace}{\theta}_{j}(s){\vert}s{\in}{\bf R}{\rbrace}$ of 
diffeomorphisms, defined by the formula
 $${\theta}_{j}(s)m=m_{j}(s), \ {\forall}m{\in}M, \ s{\in}{\bf R}.
\eqno(2.2)$$
The orbits of the ${\theta}_{j}$\rq s are termed {\it horocycles}.We note 
here that the correspondence between the group ${\theta}_{j}$ and the 
vector field $V_{j}$ is one to one since Eqs. (2.1) and (2.2) may be 
employed to define $V_{j}$ in terms of ${\theta}_{j}$ by the formula 
$$V_{j}(m)={\theta}^{\prime}(0)m \ {\forall} \ m{\in}M.\eqno(2.3)$$  
To establish consistency, we remark that this equation, together with the 
group property of ${\theta}_{j}$,  implies that
$$V_{j}\bigl({\theta}_{j}(s)m\bigr)=
{\theta}_{j}^{\prime}(0)\bigl({\theta}_{j}(s)m\bigr) =$$
$${{\partial}\over 
{\partial}t}{\theta}_{j}(t){\theta}_{j}(s)m_{{\vert}t=0}
={{\partial}\over {\partial}t}{\theta}_{j}(t+s)m_{{\vert}t=0}
={\theta}_{j}^{\prime}(s)m,\eqno(2.4)$$
as demanded by Eqs. (2.1) and (2.2).  
\vskip 0.3cm
{\bf Definition 2.1.} We term the dynamics of the model 
${\Sigma}_{cl}$ {\it hyperbolic} if the action of the differential of 
${\phi}_{t}$ on the vector fields $V_{j}$ takes the form
$$d{\phi}_{t}V_{j}(m)=V_{j}({\phi}_{t}m){\rm e}^{{\lambda}_{j}t},
\eqno(2.5)$$
where the ${\lambda}$\rq s are real numbers such that, for some positive 
integer $r$ less than $n, \ {\lambda}_{j}$ is positive for $j{\in}[1,r]$ and 
negative for $j{\in}[r+1,n]$. Thus,if $m^{\prime}$ and $m$ are 
neighbouring points of $M$ whose difference, as represented on a chart 
at $m$, is ${\sum}_{j=1}^{n}a_{j}V_{j}(m),$ the hyperbolicity 
condition signifies that
$${\phi}_{t}m-{\phi}_{t}m^{\prime}{\simeq}
{\sum}_{1}^{n}a_{j}V_{j}({\phi}_{t}m){\rm e}^{{\lambda}_{j}t}.
\eqno(2.6)$$
Hence,  defining $T_{+}(m)$ (resp. $T_{-}(m)$) to be the subspace of 
$T(m)$ spanned by the vectors $V_{j}$ for which ${\lambda}_{j}$ is 
positive (resp. negative), the hyperbolicity condition is that the action of 
${\phi}_{t}$ on neighbouring points of $M$ serves to expand their 
separation exponentially fast if their relative displacement on a chart at 
$m$ lies in $T_{+}(m)$ and contracts it if that displacement lies in 
$T_{-}(m)$. Thus the ${\lambda}$\rq s are Lyapunov exponents and, as 
some of them are positive, the hyperbolicity condition signifies that the 
flow is chaotic. The following Proposition will be proved in Appendix A
\vskip 0.3cm
{\bf Proposition 2.1.} {\it The hyperbolicity condition given by Eq. (2.5) 
is equivalent to the following one.
$${\phi}_{t}{\theta}_{j}(s){\phi}_{-t}=
{\theta}_{j}\bigl(s{\rm e}^{{\lambda}_{j}t}\bigr).\eqno(2.7)$$}
\vskip 0.3cm
{\bf  Example 1. The Arnold Cat.}\footnote{*}{This model of  
automorphisms of the torus is often so termed because of Arnold\rq s 
illustration [8] of their actions on a cat\rq s face placed in the torus.} This 
is the model $(M,{\phi},{\theta},{\mu})$, where
\vskip 0.2cm\noindent
(i) $M$ is the torus $[0,1) \ ({\rm mod}1)]^{2}$ with Euclidean netric;
\vskip 0.2cm\noindent
(ii) the time variable $t$ is discrete, its range being ${\bf Z}$, and the 
dynamical transformations are 
${\lbrace}{\phi}^{n} \ (:={\phi}_{n}){\vert}n{\in}{\bf Z}{\rbrace}$, 
where
$${\phi}=\left(\matrix{1&1\cr
                                    1&2\cr}\right);\eqno(2.8)$$  
\vskip 0.2cm\noindent
(iii) ${\mu}$ is the Lebesgue measure on the torus $M$; and
\vskip 0.2cm\noindent
(iv) denoting the eigenvectors of ${\phi}$ by $V_{1}$ and $V_{2}$ and 
their respective eigenvalues by $k_{1} \ (>1)$ and $k_{2} \ (<1),
\ {\theta}$ is the pair of one-parameter groups ${\theta}_{1}$ and 
${\theta}_{2}$ defined in terms of $V_{1}$ and $V_{2}$ by Eqs. (2.1) 
and (2.2). Thus
$${\theta}_{j}(s)m=m+V_{j}s \ \bigl({\rm mod} \ (1,1)\bigr) \ {\forall} \ 
m{\in}M, \ s{\in}{\bf R}, \ j=1,2.\eqno(2.9)$$
\vskip 0.2cm
It now follows from these definitions that the model satisfies the 
hyperbolicity condition (2.7), with ${\lambda}_{j}={\rm ln}(k_{j})$.
\vskip 0.3cm
{\bf  Example 2. Geodesic Flow on a Manifold of Negative Curvature} 
[8]. This is a model of the free dynamics of a particle on a compact region 
of the Poincare half plane 
${\tilde M}:={\lbrace}(x,y){\vert}x{\in}{\bf R},y{\in}
{\bf R}_{+}{\rbrace}$, whose metric is given by the formula
$$ds^{2}=y^{-2}(dx^{2}+dy^{2}).\eqno(2.10)$$
The points $(x,y)$ of ${\tilde M}$ will sometimes be represented by the 
complex numbers $z:=(x+iy)$.
\vskip 0.2cm
The manifold ${\tilde M}$ is equipped with the symmetry group 
$G=SL(2,{\bf R})$ [11], which acts transitively on it. The elements $g$ 
of this group are represented by two-by-two matrices with real-valued 
entries and unit determinant. Its actions on ${\tilde M}$ are given by the 
following formula. Denoting $g \ ({\in}G)$ by 
$\left(\matrix{a&b\cr
                       c&d\cr}\right),$  
$$gz={(az+b)\over (cz+d)}.\eqno(2.11)$$  
We denote by $K$ the subgroup of $G$ whose elements leave the point 
$i$ invariant. It then follows from the transitivity of $G$ that $G/K$ may 
be identified with the space ${\tilde M}$. Correspondingly, for a discrete 
co-compact non-abelian subgroup 
${\Gamma},\ {\Gamma}{\backslash}G{\slash}K$ is a compact 
manifold, ${\hat M}$, of constant negative curvature. Its unit tangent 
bundle, $T_{1}{\hat M}:=M$ may then be identified with 
${\Gamma}{\backslash}G$. We take this to be the phase space of the 
model. . 
\vskip 0.2cm
The dynamical group ${\phi}$ for the free geodesic motion of a particle 
on $M$ is given by the formula [11, 7]
$${\phi}_{t}m=m{\xi}(t),\eqno(2.12)$$
where
$${\xi}(t)=\left(\matrix{{\rm exp}(-t/2)&0\cr
                                       0&{\rm exp}(t/2)\cr}\right).
\eqno(2.13)$$
We note that the measure $d{\mu}:=y^{-2}dxdy$ is ${\phi}$-invariant. 
Further, the horocyclic actions are given by the formulae
$${\theta}_{j}(s)m=m{\xi}_{j}(s) \ {\forall} \ s{\in}{\bf R},  \   
j=1,2,\eqno(2.14)$$
where      
$${\xi}_{1}(s)=\left(\matrix{1&s\cr
                                               0&s\cr}\right) \eqno(2.15)$$
and
$${\xi}_{2}(s)=\left(\matrix{1&0\cr
                                               s&1\cr}\right) .\eqno(2.16)$$
It follows directly from these formulae that the model satisfies the 
hyperbolicity condition (2.7).
\vskip 0.5cm
\centerline {\bf  3. The Classical Operator Algebraic Picture.} 
\vskip 0.3cm 
As a first step towards a passage from the above classical picture to a 
corresponding quantum mechanical one, we now exploit the Gelfand 
isomorphism, according to which the model $(M,{\phi},{\mu})$ is 
equivalent to the $W^{\star}$ dynamic system $({\cal 
A}_{cl},{\alpha}_{cl},{\rho}_{cl})$, where 
${\cal A}_{cl}$ is the abelian $W^{\star}$ algebra of observables 
$L_{\infty}(M,d{\mu}),\ {\lbrace}{\alpha}_{cl}(t){\vert}t{\in}{\bf 
R}{\rbrace}$ is the one-parameter group of automorphisms of ${\cal 
A}_{cl}$ representing the dynamics of the model and given by the 
formula 
$$[{\alpha}_{cl}(t)A](m)=A({\phi}_{-t}m) \ {\forall} \  
A{\in}{\cal A}_{cl}, \ m{\in}M, \ t{\in}{\bf R},\eqno(3.1)$$
and ${\rho}_{cl}$ is the state on ${\cal A}_{cl}$ corresponding to the 
measure ${\mu}$, i.e.
$${\rho}_{cl}(A)={\int}Ad{\mu}.\eqno(3.2)$$ 
It follows immediately from these specifications that the 
${\phi}$-invariance of ${\mu}$ is equivalent to the ${\alpha}_{cl}$-
invariance of ${\rho}_{cl}$.
\vskip 0.2cm
Furthermore the diffeomorphism groups ${\theta}_{j}$ correspond to 
representations ${\sigma}_{j,cl}$ of ${\bf R}$ in 
${\rm Aut}({\cal A}_{cl})$, given by the formula
$$[{\sigma}_{j,cl}(s)A](m)=A\bigl({\theta}_{j}(-s)m\bigr) \ {\forall} \ 
A{\in}{\cal A}_{cl},\ m{\in}M, \ s{\in}{\bf R}.\eqno(3.3)$$
The hyperbolicity condition (2.7) is therefore equivalent to the following 
one.
$${\alpha}_{cl}(t){\sigma}_{j,cl}(s){\alpha}_{cl}(-t)=
{\sigma}_{j,cl}\bigl(s{\rm e}^{{\lambda}_{j}t}\bigr) \ {\forall} \ 
s{\in}{\bf R}, \ t \ {\in}{\bf R} \ {\rm or} \ {\bf Z}, \ j=1,. \ 
.,n.\eqno(3.4)$$
\vskip 0.5cm
\centerline {\bf 4. The Quantum Picture.}
\vskip 0.3cm
We assume that the generic quantum model corresponds to the algebraic 
picture of the classical one, but with the difference that the algebra of 
observables is non-commutative. Thus the quantum model is a triple 
$({\cal A},{\alpha},{\rho})$, where ${\cal A}$ is a $W^{\star}$-algebra, 
in general non-commutative, ${\rho}$ is a normal state on ${\cal A}$ 
and ${\lbrace}{\alpha}_{t}{\vert}t{\in}{\bf R} \ {\rm or}
{\bf Z}{\rbrace}$ is a one-parameter group of automorphisms of ${\cal 
A}$, which is continuous w.r.t. $t$ in the former case, and ${\rho}$ is a 
normal ${\alpha}$-invariant state on ${\cal A}$. Furthermore, we 
assume that the model is equipped with $n$ horocyclic actions, given by 
one-parameter groups ${\lbrace}{\sigma}_{j}(s){\vert}s{\in}{\bf R}, \ 
j=1,. \ .,n{\rbrace}$ of ${\cal A}$ whose infinitesimal generators are 
linearly independent both of one another and of that of the group 
${\alpha}$ in the case where the variable $t$ runs through ${\bf R}$. 
Accordingly, we take the hyperbolicity condition to be the natural 
generalisation of Eq. (3.4) for the possibly non-commutative case, i.e.  
$${\alpha}_{t}{\sigma}_{j}(s){\alpha}_{-t}=
{\sigma}_{j}\bigl(s{\rm e}^{{\lambda}_{j}t}\bigr) \ {\forall} \ 
s{\in}{\bf R}, \ t {\in}{\bf R} \ {\rm or} \ {\bf Z}, \ j=1,. \ .,n,
\eqno(4.1)$$
where again ${\lambda}$ is positive for $j=1,. ., r$ and negative for 
$j=r+1, . \ .,n$. 
This condition implies the following one for the duals,  
${\alpha}_{t}^{\star}$ and ${\sigma}_{j}^{\star}(s)$, of ${\alpha}_{t}$ 
and ${\sigma}_{j}(s)$, in their actions on the normal states, ${\cal 
N}({\cal A})$, on ${\cal A}$.
$${\alpha}_{-t}^{\star}{\sigma}_{j}^{\star}(s){\alpha}_{t}^{\star}=
{\sigma}_{j}^{\star}\bigl(s{\rm e}^{{\lambda}_{j}t}\bigr).\eqno(4.2)$$
We denote by ${\delta}_{j}^{\star}$ the infinitesimal generator of the 
group ${\sigma}_{j}^{\star}$, in the $w^{\star}$ topology. It follows 
from this formula that its domain, ${\cal D}({\delta}^{\star})$, is stable 
under the group ${\alpha}^{\star}$ and that, if ${\rho}_{1}$ and 
${\rho}_{2}$ are states in this domain, then
$${\Vert}{\delta}^{\star}{\alpha}_{t}^{\star}({\rho}_{1}-
{\rho}_{2}){\Vert}
={\Vert}{\delta}^{\star}({\rho}_{1}-{\rho}_{2}){\Vert}
{\rm e}^{{\lambda}_{j}t}.\eqno(4.3)$$
Thus, in the quantum context, ${\lambda}_{j}$ is a Lyapounov function 
that provideds a measure of the speed at which the evolutes of 
${\rho}_{1}$ and ${\rho}_{2}$ separate along the horocycle 
${\sigma}_{j}$. Since some of the ${\lambda}$\rq s are positive, this 
represents a chaoticity condition.
\vskip 0.2cm
We shall show, in the following Sections, that quantisation does not 
affect the hyperbolic property of the Arnold cat model, but that it destroys 
that of the geodesic flow over the manifold of negative curvature; and 
that, in general, it does not admit chaos in finite Hamiltonian systems.
\vskip 0.5cm
\centerline {\bf 5. The Quantum Arnold Cat.}  
\vskip 0.3cm
In order to quantise the classical Arnold cat model, we start by expressing 
that model in a form readily amenable to quantisation. Thus we first note 
that it follows from the definition of the classical algebra ${\cal A}_{cl}$ 
in Section 2.2 that this algebra is generated  by the sinusoidal functions 
${\lbrace}W_{cl}({\nu}){\vert}{\nu} =
({\nu}_{1},{\nu}_{2}){\in}{\bf Z}^{2}{\rbrace}$, defined by the 
formula
$$W_{cl}({\nu})[m]={\rm exp}(2{\pi}i{\nu}.m) \ {\forall} \ 
{\nu}=({\nu}_{1},{\nu}_{2}){\in}{\bf Z}^{2},\eqno(5.1)$$
where the dot denotes the Euclidean scalar product. Correspondingly, 
since ${\mu}$ is the Euclidean measure on the torus $M$, it follows from 
Eqs. (3.2) and (5.1) that
$${\rho}_{cl}\bigl(W_{cl}({\nu})\bigr)={\delta}_{{\nu},0},
\eqno(5.2)$$
where ${\delta}$ is the Kronecker delta. Moreover since, by Eq. (2.8), 
${\phi}$ is Hermitean, it follows from Eqs. (3.1) and (5.1) that
$${\alpha}_{cl}(t)W_{cl}({\nu})=
W_{cl}\bigl({\phi}_{-t}{\nu}\bigr) \ {\forall} \  t{\in}{\bf  Z}, \ 
{\nu}{\in}{\bf Z}^{2};\eqno(5.3)$$ 
while, by Eqs. (2.9), (3.3) and (5.1), the horocyclic actions for the model 
are given by the formula
$${\sigma}_{j}(s)W_{cl}({\nu})=
W_{cl}({\nu}){\rm exp}\bigl(2{\pi}i{\nu}.V_{j}s\bigr)
\ {\forall} \ s{\in}{\bf R}, \ {\nu}{\in}{\bf Z}^{2}.\eqno(5.4)$$
Thus Eqs. (5.1)-(5.4) define the classical model. One may readily check 
that they satisfy the hyperbolicity condition (3.4), bearing in mind that 
$V_{j}$ is the eigenvector of ${\phi}$ whose eigenvalue is 
${\rm exp}({\lambda}_{j})$
\vskip 0.2cm
We now quantise the classical model by basing the algebra of observables 
on Weyl operators instead of the sinusoidal function $W_{cl}$. Thus, in 
order to construct ${\cal A}$, we start with an abstract algebra of 
elements ${\lbrace}W({\nu}){\vert}{\nu}{\in}{\bf Z}^{2}{\rbrace}$ 
which satisfy the Weyl condition that
$$W({\nu})W({\nu}^{\prime})=W({\nu}+{\nu}^{\prime})
{\rm exp}\bigl(i{\gamma}{\kappa}({\nu},{\nu}^{\prime})\bigr),
\eqno(5.5)$$
where ${\kappa}$ is the simplectic form defined by the formula
$${\kappa}({\nu},{\nu}^{\prime})={\nu}_{1}{\nu}_{2}^{\prime}
-{\nu}_{2}{\nu}_{1}^{\prime}\eqno(5.6)$$
and ${\gamma}$ is a constant that plays the role of that of Planck. Thus 
the algebra ${\cal A}_{0}$ of the polynomials in the $W({\nu})$\rq s 
comprises just the linear combinations of them. We define ${\rho}$ to be 
the positive normalised linear form on this algebra given by the precise 
analogue of the classical state ${\rho}_{cl}$, as given by Eq. (5.2), i.e.
$${\rho}\bigl(W({\nu})\bigr)={\delta}_{{\nu},0}.\eqno(5.7)$$
We define the algebra of observables, ${\cal A}$, to be the strong closure 
of the GNS repreesentation of ${\cal A}_{0}$ in the state ${\rho}$ 
defined by this last equation. We then define the dynamical and 
horocyclic automorphisms, ${\alpha}$ and ${\sigma}_{j}$, by the 
canonical counterparts of the classical ones of Eqs. (5.3) and (5.4). Thus
$${\alpha}(t)W({\nu})=
W\bigl({\phi}_{-t}{\nu}\bigr) \ {\forall} \  t{\in}{\bf  Z}, \ 
{\nu}{\in}{\bf Z}^{2};\eqno(5.8)$$ 
and
$${\sigma}_{j}(s)W({\nu})=
W({\nu}){\rm exp}\bigl(2{\pi}i{\nu}.V_{j}s\bigr)
\ {\forall} \ s{\in}{\bf R}, \ {\nu}{\in}{\bf Z}^{2}.\eqno(5.9)$$
It follows from the last two formulae that the model satisfies the 
hyperbolicity condition (4.1). Thus we have established the following 
Proposition.
\vskip 0.3cm
{\bf Proposition 5.1.} {\it The chaoticity of the flow of the Arnold cat 
model survives quantisation.}
\vskip 0.5cm
\centerline {\bf 6. Quantum Geodesic Flow on a Compact Manifold of 
Negative Curvature.} 
\vskip 0.3cm
The model we now consider is the quantised version of that of Example 2 
in Section 2, and it may be described as follows [11, 7]. Its $W^{\star}$-
algebra of observables is ${\cal B}({\cal H})$, the set of bounded 
operators in the Hilbert space ${\cal H}:= L_{2}({\hat M},d{\mu})$, 
where the measure $d{\mu}$ is defined following Eq. (2.13). The state 
space of the model comprises the normal states of ${\cal A}$ and its 
Hamiltonian is $-{\Delta}$, where ${\Delta}$ is the Laplace-Beltrami 
operator for the manifold ${\hat M}$. The dynamical automorphisms of 
the model are thus given by the formula
$${\alpha}_{t}A={\rm exp}(-i{\Delta}t)A{\rm exp}(i{\Delta}t) \ 
{\forall} \  A{\in}{\cal A}, \ t{\in}{\bf R}.\eqno(6.1)$$.
Moreover, the spectrum of ${\Delta}$ is discrete [12]. We denote by 
${\lbrace}f_{k}{\vert}k{\in}{\bf N}{\rbrace}$ a complete orthonormal 
set of eigenvectors of this operator and by ${\lbrace}e_{k}{\rbrace}$ the 
corresponding set of its eigenvalues. We then define the operarors 
$F_{kl}$, with $k,l{\in}{\bf N}$, by the equation
$$F_{kl}f_{i}={\delta}_{li}f_{k} \ {\forall}k,l.i{\in}{\bf N}.
\eqno(6.2)$$
It follows now from Eqs. (6.1) and (6.2) that
$${\alpha}_{t}F_{kl}={\rm exp}(i{\omega}_{kl}t)F_{kl} \ {\forall} \ 
t{\in}
{\bf R}, \ k,l{\in}{\bf N},\eqno(6.3)$$
where
$${\omega}_{kl}=e_{l}-e_{k}.\eqno(6.4)$$
We denote by ${\cal L}(F)$ the set of finite linear combinations of the 
$F_{kl}$\rq s. It follows from this definition that ${\cal L}(F)$ is closed 
with respect to involution and binary addition and multiplication. It is 
therefore a $^{\star}$-algebra, and it follows from our specifications that 
its strong closure is ${\cal A}$. 
\vskip 0.3cm
{\bf Proposition 6.1.} {\it Under the above assumptions, the quantum 
geodesic flow on the manifold cannot be hyperbolic.}
\vskip 0.3cm
We base the proof of this Proposition on Lemmas (6.2) and (6.3) below.
\vskip 0.3cm
{\bf Lemma 6.2.} {\it Assume that the model satisfies the hyperbolicity 
condition with respect to horocyclic automorphisms ${\sigma}({\bf R})$. 
Then it follows from the discretenes of the spectrum of ${\Delta}$ that 
any normal stationary state ${\rho}$ of the model is ${\sigma}$-
invariant.}
\vskip 0.3cm
Assuming the result of this lemma, we denote the GNS triple for the state 
${\rho}$ by $({\cal H}_{\rho},{\pi}_{\rho},{\Phi}_{\rho})$ and define  
$U_{\rho}$ and $V_{\rho}$ to be the continuous unitary representations 
of ${\bf R}$ in ${\cal H}_{\rho}$ that implement the automorphisms 
${\alpha}_{t}$ and ${\sigma}(s)$, respectively, according to the 
standard prescription
$$U_{\rho}(t){\pi}_{\rho}(A){\Phi}_{\rho}=
{\pi}_{\rho}({\alpha}_{t}A){\Phi}_{\rho}\eqno(6.5)$$ 
and
$$V_{\rho}(s){\pi}_{\rho}(A){\Phi}_{\rho}=
{\pi}_{\rho}\bigl({\sigma}(s)A\bigr){\Phi}_{\rho}.\eqno(6.6)$$
Hence, by the cyclicity of ${\Phi}_{\rho}$ and the hyperbolicity 
condition (4.1), as applied to the horocycle ${\sigma}$, that
$$U_{\rho}(t)V_{\rho}(s)U_{\rho}(-t)=
V_{\rho}(se^{{\lambda}t}) \ {\forall} \ s,t{\in}
{\bf R}\eqno(6.7)$$
We define $H_{\rho}$ to be the Hamiltonian operator in the GNS space, 
${\cal H}_{\rho}$, according to the formula 
$U_{\rho}(t)={\rm exp}(iH_{\rho}t)$. 
\vskip 0.3cm
{\bf Lemma 6.3.} {\it  Under the assumptions of Lemma 6.2 and with the 
subsequent definitions, the formula (6.7) implies that the spectrum of 
$H_{\rho}$ is ${\bf R}$.}
\vskip 0.3cm
{\bf Proof of Prop. 6.1 assuming Lemmas 6.2 and 6.3.} Our strategy here 
is to infer from Lemma 6.2 that the the assumption of hyperbolicity 
implies that the spectrum of $H_{\rho}$ is discrete. Since this conflicts 
with Lemma 6.3, we conclude that that assumption is invalid.  
\vskip 0.2cm
We start by noting that, by Eqs. (6.3) and (6.5), 
$$U_{\rho}(t){\pi}_{\rho}(F_{kl}){\Phi}_{\rho}=
{\pi}_{\rho}(F_{kl}){\Phi}_{\rho}{\rm exp}(i{\omega}_{kl}t).
\eqno(6.8)$$
Since ${\rho}$ is a normal stationary state of the model, it follows from 
the definition of the vectors $f_{k}$ that ${\rho}$  corresponds to a 
density matrix of the form ${\sum}_{r{\in}{\bf  N}}w_{r}P_{r}$, 
where the $w_{r}$\rq s are non-negative numbers whose sum is unity 
and $P_{r} \ (=F_{rr})$ is the projection operator for the vector $f_{r}$. 
Hence
$${\langle}{\pi}_{\rho}(A){\Phi}_{\rho},{\pi}_{\rho}(B)
{\Phi}_{\rho}{\rangle}={\langle}{\rho};(A^{\star}B{\rangle}
={\sum}_{r{\in}{\bf N}}w_{r}(f_{r},A^{\star}Bf_{r}) \ {\forall} \ 
A,B{\in}{\cal A}.\eqno(6.9)$$
It follows from this formula and Eq. (6.2) that
$${\langle}{\pi}_{\rho}(F_{kl}){\Phi}_{\rho},
{\pi}_{\rho}(F_{k^{\prime}l^{\prime}}{\Phi}_{\rho}{\rangle}=
w_{l}{\delta}_{kk^{\prime}}{\delta}_{ll^{\prime}}.\eqno(6.10)$$
Therefore, defining $D:={\lbrace}(k,l){\in}
{\bf N}^{2};w_{l}{\neq}0{\rbrace}$ and
$${\Psi}_{kl}=
w_{l}^{-1/2}{\pi}_{\rho}(F_{kl}){\Phi}_{\rho} \ {\forall} \ 
(k,l){\in}D,\eqno(6.11)$$
the set of vectors ${\lbrace}{\Psi}_{kl}{\vert}(k,l){\in}D{\rbrace}$ is 
orthonormal. It is also complete for the  following reasons. By the 
definition (6.2) of the operators $F_{kl}$, the algebra ${\cal A}$ consists 
of linear combinations of these operators. Therefore, by the normality of 
the representation ${\pi}_{\rho}$, the algebra ${\pi}_{\rho}({\cal A})$ 
consists of linear combinations of  the operators ${\pi}_{\rho}(F_{kl})$. 
Hence by Eq. (6.11) and the cyclicity of ${\Pi}_{\rho}$ with respect to 
that algebra, the set ${\lbrace}{\Psi}_{kl}{\vert}(k,l){\in}D{\rbrace}$ of 
orthonormal vectors in ${\cal H}_{\rho}$ is complete. 
\vskip 0.2cm
Now, by Eqs. (6.8) and (6.11),
$$U_{\rho}(t){\Psi}_{kl}={\Psi}_{kl}{\rm exp}(i{\omega}_{kl}t) \  
{\forall} \ (k,l){\in}D.$$
and consequently, since 
${\lbrace}{\Psi}_{kl}{\vert}(k,l){\in}D{\rbrace}$ is an orthonormal 
basis in ${\cal H}_{\rho}$, 
$$U_{\rho}(t)={\sum}_{(k,l){\in}D}
{\cal P}_{kl}{\rm exp}(i{\omega}_{kl}t),$$
where ${\cal P}_{kl}$ is the projector for ${\Psi}_{kl}$. Hence
$$H_{\rho}={\sum}_{(k,l){\in}D}{\omega}_{kl}{\cal P}_{kl},
\eqno(6.12)$$
and therefore the spectrum of $H_{\rho}$ comprises the discrete set 
${\omega}_{kl}{\vert}(k,l){\in}D$. As this conflicts with Lemma 6.3, 
which was based on the assumption of a hyperbolic flow, we conclude 
that the model does not support such a flow.
\vskip 0.3cm
{\bf Proof of Lemma 6.2.} By the hyperbolicity condition (4.1), as 
applied to the horocycle ${\sigma}$,
$${\langle}{\rho};{\alpha}_{t}{\sigma}(s){\alpha}_{-t}F_{kl}{\rangle}
={\langle}{\rho};{\sigma}\bigl(se^{{\lambda}t})F_{kl}{\rangle}.
\eqno(6.8)$$
By Eq. (6.3) and the stationarity of ${\rho}$, the l.h.s. of this equation is 
equal to ${\langle}{\rho}:{\sigma}(s)F_{kl}{\rangle}
{\rm exp}(i{\omega}_{lk}t)$. On the other hand, in the limit where 
${\lambda}t{\rightarrow}-{\infty}$, it follows by continuity that the 
r.h.s. of Eq. (6.8) reduces to ${\langle}{\rho};F_{kl}{\rangle}$. 
Compatibility of these expressions for the two sides of Eq. (6.8) implies 
that ${\langle}{\rho};{\sigma}(s)F_{kl}{\rangle}$ and 
${\langle}{\rho};F_{kl}{\rangle}$ are equal to one another if  
${\omega}_{kl}=0$ and are both zero if ${\omega}_{kl}{\neq}0$. 
Hence they are equal in all cases. In view of the normality of ${\rho}$ 
and the strong density of ${\cal L}(F)$, this result  implies that ${\rho}$ 
is ${\sigma}$-invariant.
\vskip 0.3cm
 {\bf Proof of Lemma 6.3.} This is achieved in Ref. [7] on the basis of a 
version of Mackey\rq s imprimitivity theorem.
\vskip 0.5cm
\centerline {\bf 7. Generic Non-Hyperbolic Flow of Finite Quantum 
Hamiltonian Systems.} 
\vskip 0.3cmThe generic model of a finite quantum Hamiltonian system 
is not quite the same as the model presented in Section 4. Specifically it 
consists of a triple $({\cal A},{\alpha},{\cal N})$ [10, 13], where ${\cal 
A}$ is the $W^{\star}$-algebra of bounded operators in a separable 
Hilbert space ${\cal H}, \ {\cal N}$ is the set of normal states on ${\cal 
A}$ corresponding to the density matrices in ${\cal H}$, and ${\alpha}$ 
is a representation of ${\bf R}$ in the automorphisms of ${\cal A}$ 
implemened by a unitary group  whose infinitesimal generator is $i$ 
times a self-adjoint operator $H$. Thus
$${\alpha}_{t}A={\rm exp}(iHt/{\hbar})A{\rm exp}(-iHt/{\hbar}) \ 
{\forall} \ A{\in}{\cal A}, \ t{\in}{\bf R}.\eqno(7.1)$$
Here $H$ is the Hamiltonian of the model. In general, it is the sum of the 
kinetic and potential energies of its constituent particles and its spectrum 
is discrete. Note that these specifications do not include the assumption of 
a hyperbolicity assumption such as given by Eq. 4.1. In fact, the 
following proposition establishes the contrary of that assumption for this 
model. 
\vskip 0.3cm 
{\bf Proposition 7.1.} {\it Finite quantum Hamiltonian systems, as 
defined above, cannot support hyperbolic flows.}
\vskip 0.3cm
{\bf Proof.} This follows immediately from the discreteness of the 
spectrum of $H$ by the same argument that led from Lemmas 6.2 and 6.3 
to Prop. 6.1.
\vskip 0.5cm
\centerline {\bf 8. Conclusions}
\vskip 0.3cm
The general picture of quantum hyperbolic flows, presented in Section 4, 
is the natural analogue of its algebraically cast classical counterpart and 
exhibits the chaotic property represented by Eq. (4.3). Moreover, this 
picture is realised by the quantum Arnold cat model. On the other hand,  
finite quantum Hamiltonian systems, including the geodesic flow over a 
compact manifold of constant negative curvature, do not support  
hyperbolic flows. This accords with a vast body of work on models for 
which chaos in classical systems is suppressed by quantisation [5, 6]. 
Since, in those works, the classical chaos leaves its mark on the resultant 
quantum system in the form of certain scars on its eigenstates, we expect 
that this is also the case for the quantum Hamiltonian models treated here.   
\vskip 0.5cm 
\centerline {\bf Appendix A: Proof of Proposition 2.1.}
\vskip 0.3cm
In order to derive Eq. (2.7) from Eq. (2.5), we start by defining
$${\tilde m}_{j,t}(s)={\phi}_{t}{\theta}_{j}
\bigl({\rm exp}(-{\lambda}_{j}t)s\bigr){\phi}_{-t}m \ {\forall} 
s,t{\in}{\bf R}, \ m{\in}M, \ j=1,. \ .,n\eqno(A.1)$$
and inferring from this formula that, for fixed $t$ and $j$,
$${\tilde m}_{j,t}^{\prime}(s)={\rm exp}(-{\lambda}_{j}t)
d{\phi}_{t}{\theta}_{j,t}^{\prime}
\bigl({\rm exp}(-{\lambda}_{j}t)s\bigr){\phi}_{-t}m\eqno(A.2)$$
Hence, by Eq. (2.1),
$${\tilde m}_{j,t}^{\prime}(s)={\rm exp}(-{\lambda}_{j}t)
d{\phi}_{t}V\bigl({\theta}_{j,t}
\bigl({\rm exp}(-{\lambda}_{j}t)s\bigr){\phi}_{-t}m.$$
 and therefore, by Eqs. (2.5) and (A.1),
$${\tilde m}_{j,t}^{\prime}(s)=V\bigl({\tilde m}_{j,t}(s)\bigr),
\eqno(A.3)$$
which signifies that ${\tilde m}_{j,t}(s)$ is the unique solution of Eq. 
(2.4), i.e. that ${\tilde m}_{j,t}(s)={\theta}_{j}(s)m.$ In view of Eq. 
(A.1), this implies that
$${\phi}_{t}{\theta}_{j}
\bigl({\rm exp}(-{\lambda}_{j}t)s\bigr){\phi}_{-t}={\theta}_{j}(s), 
\ {\forall} \ s,t{\in}{\bf R}, \ m{\in}M,\ j=1,. \ .n,$$ 
which is equivalent to Eq. (2.7). 
\vskip 0.2cm
Conversely, in order to derive Eq. (2.5) from Eq. (2.7), we note that, in 
view of the formula (A.1), the latter equation signifies that 
${\tilde m}_{j,t}(s)=m_{j}(s)$. Hence, by Eq. (2.1),
$${\tilde m}_{j,t}^{\prime}(s)=
V\bigl({\tilde m}_{j,t}(s)\bigr).\eqno(A.4)$$
Furthermore, by Eq. (A.1), the l.h.s. of this formula is equal to
$${{\partial}\over {\partial}s}{\phi}_{t}{\theta}
\bigl({\rm exp}(-{\lambda}_{j}t)s\bigr){\phi}_{-t}m=
{\rm exp}(-{\lambda}_{j}t)d{\phi}_{t}{\theta}_{j}^{\prime}
\bigl({\rm exp}(-{\lambda}_{j}t)s\bigr){\phi}_{-t}m,$$
which, by Eq. (2.1), is equal to
$${\rm exp}(-{\lambda}_{j}t)d{\phi}_{t}V\bigl({\theta}_{j}
({\rm exp}(-{\lambda}_{j}t)s\bigr){\phi}_{-t}m.$$
Hence, by Eq. (A.1),  Eq. (A.4) reduces to the form
$$d{\phi}_{t}V\bigl({\theta}_{j}\bigl({\rm exp}(-{\lambda}_{j}t)s\bigr)
{\phi}_{-t}m\bigr)={\rm exp}({\lambda}_{j}t)
V\bigl({\tilde m}_{j,t}(s)\bigr),$$
i.e., by Eq. (A.1),
$$ d{\phi}_{t}V\bigl({\phi}_{-t}{\tilde m}_{j,t}(s)\bigr) =
{\rm exp}({\lambda}_{j}t)V\bigl({\tilde m}_{j,t}(s)\bigr).$$
Thus, putting
$${\hat m}={\phi}_{-t}{\tilde m}_{j,t}(s).\eqno(A.5)$$
$$d{\phi}_{t}V({\hat m})=V({\phi}_{t}{\hat m}).$$
Since, by Eqs. (A.1) and (A.5), the correspondence between $m$ and 
${\hat m}$ is one-to-one, this last equation is equivalent to Eq. (2.5).
\vskip 0.5cm
\centerline {\bf References}
\vskip 0.3cm\noindent
[1] D. V. Anosov: Proc. Inst. Steklov {\bf 90}, 1 (1967)
\vskip 0.2cm\noindent
[2] I. Farquhar: {\it Ergodic theory in Statistical Mechanics}, Wiley, New 
York, 1964
\vskip 0.2cm\noindent
[3] G. L. Sewell: Pp. 511-538 of {\it Lectures in Theoretical Physics}, 
Vol.XIVB, Ed. W. E. Brittin, Colorado University Press, 1973
\vskip 0.2cm\noindent
 [4] G. G. Emch and C. Liu: {\it The Logic of Thermostatistical Physics}, 
Springer, Berlin, Heidelberg, New York, 2002
\vskip 0.2cm\noindent
[5] M. Gutzwiller: {\it Chaos in Classical and Quantum Mechanics}, 
Springer, Berlin, Heidelberg, London, 1990
\vskip 0.2cm\noindent
[6] F. Haake: {\it Signatures of Quantum Chaos}, Springer, Berlin, 
Heidelberg, New York, 1992
\vskip 0.2cm\noindent
[7] G. G. Emch, H. Narnhofer, W. Thirring and G. L. Sewell: J. Math. 
Phys. {\bf 35}, 5582 (1994)
\vskip 0.2cm\noindent
[8] V. I. Arnold and A. Avez: {\it Ergodic Problems of Classical 
Mechanics}, Benjamin, New York, 1969
\vskip 0.2cm\noindent
 [9] G. G. Emch: {\it Algebraic Methods in Statistical Mechanics and 
Quantum Field Theory}, Wiley, New York, 1972
\vskip 0.2cm\noindent
[10] G. L. Sewell: {\it Quantum Mechanics and its Emergent 
Macrophysics}, Princeton University Press, Princeton, 2002
\vskip 0.2cm\noindent
[11] G. G. Emch: J. Math. Phys. {\bf 23}, 1785, (1982)
\vskip 0.2cm\noindent
[12] M.Berger, B. Granduchon and E. Mazed: {\it Le Spectre d\rq une 
Variete Riemannienne}, Springer Lecture Notes in Mathematics 
{\bf 194}, 1971
\vskip 0.2cm\noindent
[13] J. Von Neumann: {\it Mathematical Foundations of Quantum 
Mechanics}, Princeton University Press, Princeton, 1955.

\end